\documentclass[aps]{revtex4}
\usepackage{graphicx}
\usepackage{amsmath}
\usepackage{amssymb}
\newcommand{\bb}{\begin{eqnarray}}
\newcommand{\ee}{\end{eqnarray}}

\begin{document}

\title{Entropy of charged dilaton-axion black hole}
\author{Tanwi Ghosh\footnote{E-mail: tanwi.ghosh@yahoo.co.in} and Soumitra SenGupta\footnote{E-mail: tpssg@iacs.res.in}}
\affiliation{Department of Theoretical Physics, Indian Association for the
Cultivation of Science,\\
Jadavpur, Calcutta - 700 032, India}

\begin{abstract}
Using brick wall method the entropy of charged dilaton-axion black hole is determined for both
asymptotically flat and non-flat cases. The 
entropy turns out to be  proportional to the horizon area of the black hole confirming  
the Beckenstien, Hawking area-entropy formula for black holes.  
The leading order logarithmic corrections to the entropy are also derived for 
such black holes.
\end{abstract}
\maketitle

{{\Large {\bf 1. Introduction :}}}\\

Ever since Beckenstein-Hawking interpreted the black hole entropy in terms of  
the area of the black hole horizon \cite{Bekenstein}, there have been numerous works in this direction.
One of the main focus of such works has been to explain the statistical
origin of black hole entropy. It is well known that the statistical entropy arises
from the microscopic quantum states which are macroscopically indistinguishable.
To identify such states 
t'Hooft\cite{t hooft} argued that the origin of black hole entropy can be explained from states
of quantum fields propagating outside the black hole horizon. To eliminate
divergences which appear in the counting of states just on the horizon, brick wall
method was used. Subsequently the efficacy of the brick wall method in calculating various thermodynamic properties of black holes 
has been extensively shown in different contexts \cite{Suss,Uglum,Cai,Winstanley,Kim,Zhong,Cognola,Kim2,Demers,Ghosh,Hawking,
TG,Solo,ELC,SS,WYY,ZZ,MKK,RS,ZW,LH,YO,WY,WE,SM,RZL,RL,YW,X,LI,K,Re,We,EL,LZ}.
In this work we  consider charged black hole in generalised dilaton-axion
gravity \cite{Sur},
and apply the brick wall method to verify the Beckenstein-Hawking area law for black hole entropy. We further 
investigate the possible  leading order correction to the entropy obtained from the brick wall calculation.
Our calculation yields a logarithmic correction to the area law for these class of black holes which is known to be
a generic feature for other black holes also.  

We begin with a generalised action for two scalar fields coupled with
Einstein-Hilbert-Maxwell field in  four dimension:
\begin{eqnarray}
S=\frac{1}{2\kappa}\int d^4
 x\sqrt{-g}[R-\frac{1}{2}\partial_{\mu}\varphi\partial^{\mu}\varphi
-\frac{\omega(\varphi)}{2}\partial_{\mu}\zeta\partial^{\mu}\zeta-\alpha(\varphi,\zeta)F^2
-\beta(\varphi,\zeta)F_{\mu\nu}*F^{\mu\nu}]
\end{eqnarray}
where $\kappa=8\pi G$,R is the curvature scalar,$F_{\mu\nu}$ is the
 Maxwell field,${\varphi,
\zeta}$ are two massless scalar/pseudo-scalar fields which couple to the Maxwell field with arbitrary 
functional dependence given by $\alpha(\varphi,\zeta) $ and $\beta(\varphi,\zeta)$.
Motivation for choosing such action is rooted  in string low energy effective field theory where the two massless scalar/pseudo-scalar fields
namely dilaton/axion indeed couple to the Maxwell field in a specific way. It has been shown that interesting black hole
solutions exist in such scenario. The above action is a natural generalization to look for a much wider class of 
black hole solutions \cite{Sur}.     

Starting from the above action the
solutions for the metric for the asymptotically flat and non-flat cases are obtained \cite{Sur} as,\\

{\bf Asymptotically flat metric :}\\
\begin{eqnarray}
ds^2= -\frac{(r-r_+)(r-r_-)}{(r-r_o)^{2-2n}(r+r_o)^{2n}}dt^2
 +\frac{(r-r_o)^{2-2n}(r+r_o)^{2n}}{(r-r_+)(r-r_-)}dr^2+\frac{(r+r_o)^{2n}}{(r-r_o)^{(2n-2)}}
 d\Omega^2       
\end{eqnarray}\\
{\bf Asymptotically non-flat metric :}\\
\begin{eqnarray}
ds^2= -\frac{(r-r_+)(r-r_-)}{r^2(\frac{2r_o}{r})^{2n}}dt^2
 +\frac{r^2(\frac{2r_o}{r})^{2n}}{(r-r_+)(r-r_-)}dr^2+
 r^2(\frac{2r_o}{r})^{2n}d\Omega^2       
\end{eqnarray}\\

Various parameters for the asymptotically flat case are given as\cite{Sur},\\

$r_{\pm} = m_o \pm\sqrt{m_o^2+r_o^2-\frac{1}{8}(\frac{K_1}{n}+\frac{K_2}{1-n})}$ ; $r_o=\frac{1}{16m_o}(\frac{K_1}{n}-\frac{K_2}{1-n})$
 
$m_o= m-(2n-1)r_o$ ; $K_1 = 4n[4r_o^2 +2kr_o(r_++r_-)+k^2r_+r_-]$ ;  $K_2=4(1-n)r_+r_-$;  $o<n<1$ 
and $ m = \frac{1}{16r_o}(\frac{K_1}{n}-\frac{K_2}{1-n})+(2n-1)r_o$, where $m$ is the mass of the black hole and k=1 for asymptotically flat case\\.

Similarly for asymptotically non-flat case\\
$ r_{\pm}= (\frac{1}{(1-n)})[m\pm\sqrt{m^2-(1-n)\frac{K_2}{4}}]$\\
and $ K_1  = 4n[4r_o^2 +2kr_o(r_++r_-)]$ ; $K_2=4(1-n)r_+r_-$;k=0 for asymptotically non-flat case.

We now calculate the entropy of such a charged dilaton-axion black hole 
for both asymptotically flat and non flat cases using the brick wall method.
Considering a scalar field in the background of such a black hole, the 
partition function for the system can be written as \cite{Ghosh}:
\begin{eqnarray}
Z=e^{-S_1[g_{cl},\varphi_{cl},F_{cl},\zeta_{cl}]}\int[d\phi]e^{-S_2[g_{cl},\phi]}
\end{eqnarray}
Here  $\phi$ is a scalar field in the background of 
the dilaton-axion black hole metric. In deriving this partition function
quantum fluctuations of the background metric,the electromagnetic field and the
dilaton-axion fields are not considered and these fields are treated as background classical fields in such
semi-classical approximation.\\

{{\Large{\bf 2. The brick wall model and calculation of entropy:}}}\\

While calculating the number of energy levels 
that a particle can occupy near the black hole one encounters a
divergence as one hits the black hole horizon. This is the source of the  usual divergence in the 
expression for free energy of the scalar field. To avoid this 
divergence, the particle wave function extremely close to the horizon 
must be modified by gravitational interaction
between the ingoing and outgoing particles.
Following 't Hooft [15], we assume a cut-off  on the wave function 
$\phi(x)$ just outside the horizon.
This is given as,
\begin{eqnarray}
\phi(x)=0
\end{eqnarray}
at
\begin{eqnarray}
 r=r_h +\epsilon
\end{eqnarray}
where $\epsilon$ is a small, positive quantity and signifies an 
ultraviolet cut-off. 
There is also an infrared cut-off
\begin{eqnarray}
\phi(x)=0
\end{eqnarray} 
at 
\begin{eqnarray}
r=L
\end{eqnarray}
with $L>>r_h$.\\

{{\Large{\bf 2.1. Asymptotically flat dilaton-axion black hole:}}}\\

The wave equation for a scalar field (boson field) in an
asymptotically flat dilaton-axion coupled space-time is given as,
\begin{eqnarray}
\partial_{\mu}(\sqrt{-g}g^{\mu\nu}\partial_{\nu}\phi)-m^2\phi = o
\end{eqnarray}
Expanding the scalar field in spherical harmonics,
\begin{eqnarray}
\phi = e^{-iEt}f_{El}Y_{lm_l}
\end{eqnarray}
Using Equ.(2) and Equ.(9) in Equ.(8),the equation for the scalar field in the asymptotically flat dilaton-axion black hole background becomes,
\begin{eqnarray}
\nonumber\frac{E^2f_{El}}{\frac{(r-r_+)(r-r_-)}{(r-r_0)^{2-2n}(r+r_0)^{2n}}}+\frac{1}{\frac{(r+r_0)^{2n}}{(r-r_0)^{2n-2}}}
\nonumber\frac{\partial}{\partial
 r}[\frac{(r+r_0)^{2n}}{(r-r_0)^{2n-2}}\frac{(r-r_+)(r-r_-)}{(r-r_0)^{2-2n}(r+r_0)^{2n}}\frac{\partial
 f_{El}}{\partial r}]
-[\frac{l(l+1)}{\frac{(r+r_0)^{2n}}{(r-r_0)^{2n-2}}}+m^2]f_{El}= 0
\end{eqnarray} \\

This leads to an $r$ dependent radial wave number:
\begin{eqnarray}
k^2(r,l,E)=\frac{[E^2-\frac{(r-r_+)(r-r_-)}{(r-r_0)^{2-2n}(r+r_0)^{2n}}(\frac{l(l+1)}{\frac{(r+r_0)^2n}{(r-r_0)^{2n-2}}}+m^2)]}
{(\frac{(r-r_+)(r-r_-)}{(r-r_0)^{2-2n}(r+r_0)^{2n}})^2}
\end{eqnarray}\\

Following semi-classical quantisation rule, the number of radial modes $n$ can be written as 
\begin{eqnarray}
 n_r\pi = \int_{r_h+\epsilon}^L dr k(r,l,E)
\end{eqnarray} \\

To determine the thermodynamic properties of the system we first consider the
free energy of a thermal ensemble of scalar particles with an inverse
temperature $\beta$ as follows ( up to logarithmic corrections):

\begin{eqnarray}
-\beta F_{boson} &=&\sum_{n_r,l,m_l}log(1-e^{-\beta E})
\nonumber \approx\int dl(2l+1)\int dn_r log(1-e^{-\beta E})\\
\end{eqnarray}
\begin{eqnarray}
\nonumber && =-\int dl (2l+1)\int d(\beta E)\frac{n_r}{(e^{\beta 
E}-1)}\\
\nonumber &&  =-\frac{\beta}{\pi}\int dl (2l+1)\int \frac{dE}{(e^{\beta
E}-1)}
\nonumber\int dr A(r)  \\
\nonumber &&=-\frac{2\beta}{3\pi}\int_{r_h+\epsilon}^L
 dr\frac{\frac{(r+r_0)^2n}{(r-r_0)^{2n-2}}}
               {\Delta^2} \int_0^\infty dE B(r)\\   
\end{eqnarray}\\
where 
\begin{eqnarray}
A(r) = 
\frac{[E^2-\frac{(r-r_+)(r-r_-)}{(r-r_0)^{2-2n}(r+r_0)^{2n}}(\frac{l(l+1)}{\frac{(r+r_0)^2n}
{(r-r_0)^{2n-2}}}+m^2)]^\frac{1}{2}}
{\frac{(r-r_+)(r-r_-)}{(r-r_0)^{2-2n}(r+r_0)^{2n}}}
\end{eqnarray}
\begin{eqnarray}
B(r)= 
=\frac{[E^2-\frac{(r-r_+)(r-r_-)}{(r+r_0)^{2n}(r-r_0)^{2-2n}}m^2]^{3/2}}{(e^{\beta
 E}-1)}
\end{eqnarray}
and 
\begin{eqnarray}
\Delta = \frac{(r-r_+)(r-r_-)}{(r-r_0)^{2-2n}(r+r_0)^{2n}}
\end{eqnarray}
For a nonextremal black hole we expand $\Delta$ about the event horizon,
\begin{eqnarray}
\Delta = \Delta_h'(r-r_h)+\frac{1}{2}\Delta_h''(r-r_h)^2+O(r-r_h)^3
\end{eqnarray}
The other quantities in equation (14) can be expanded.Here the limits of integration is taken for values where the square root is real. 
While the integration over $l$ has been done accurately, the E integral has been evaluated only approximately.\\
\begin{eqnarray}
\beta F_{boson}&\approx&
\nonumber\frac{-2\beta}{3\pi}\frac{\frac{(r_++r_0)^{2n}}{(r_+-r_0)^{(2n-2)}}}{\Delta_h'^2}[\frac{1}{\epsilon}\frac{\pi^4}{15\beta^4}\\
&&-(\frac{2}{\frac{(r_++r_0)^n}{(r_+-r_0)^{(n-1)}}}-\frac{\Delta_h''}{\Delta_h'})\frac{\pi^4}{15\beta^4} ln(\frac {L}{\epsilon})]
\end{eqnarray}
where
\begin{eqnarray}
\Delta_h'=(r_h-r_-)(r_h-r_0)^{(2n-2)}(r_h+r_0)^{-2n} 
\end{eqnarray}
and 
\begin{eqnarray}
\Delta_h''&=& (r_h-r_0)^{(2n-2)}(r_h+r_0)^{-2n}+(2n-2)(r_h-r_-)(r_h-r_0)^{(2n-3)}(r_h+r_0)^{-2n} \nonumber\\
&-& 2n(r_h-r_-)(r_h-r_0)^{(2n-2)}(r_h+r_0)^{(-2n-1)}+ (r_h-r_0)^{(2n-2)}(r_h+r_0)^{-2n} \nonumber\\
&+& (2n-2)(r_h-r_-)(r_h-r_0)^{(2n-3)}(r_h+r_0)^{-2n} -2n(r_h-r_-)(r_h-r_0)^{(2n-2)}(r_h+r_0)^{-2n-1} 
\end{eqnarray}
For $r_h=r_+$ the expression for entropy can now be obtained from the following
  formula.\\
\begin{eqnarray}
S_{boson} = \beta^2\frac{\partial F}{\partial \beta}
=\frac{1}{\epsilon}\frac{\frac{(r_++r_0)^{2n}}{(r_+-r_0)^{(2n-2)}}}{360}\Delta_h'-[\frac{\frac{(r_++r_0)^{2n}}{(r_+-r_0)^{(2n-2)}}}{360}(\frac{2\Delta_h'}{\frac{(r_++r_0)^n}{(r_+-r_0)^{(n-1)}}}-\Delta_h'')]ln(\frac {L}{\epsilon})
\end{eqnarray}
$\beta$  can be calculated from the metric ( see equ.2) as follows,
\begin{eqnarray}
\beta = \frac{\partial }{\partial r}[\frac{(r-r_+)(r-r_-)}{(r-r_o)^{2-2n}(r+r_o)^{2n}}]_{r=r_+}
      =\frac{(r_o+r_+)^{2n}(r_+-r_o)^{2-2n}}{(r_+-r_-)}
\end{eqnarray}
The  distance of the brick wall from the horizon is related to the ultraviolet cutoff $\epsilon$ as
\begin{eqnarray}
\nonumber p =\int_{r_+}^{r_{+}+\epsilon}ds =
 \int_{r_+}^{r_{+}+\epsilon}dr[\frac{(r-r_+)(r-r_-)}{(r+r_o)^{2n}(r-r_o)^{2-2n}}]^{-\frac{1}{2}}\\
\epsilon = \frac{(r_+-r_-)(r_+-r_o)^{2(n-1)}p^2}{4(r_o+r_+)^{2n}}
\end{eqnarray}
Substituting the expression for $\beta$ and $\epsilon$ in the expression for entropy ( in the leading order ) we have
\begin{eqnarray}
S_{boson}=\frac{\frac{(r_++r_0)^{2n}}{(r_+-r_0)^{(2n-2)}}}{90p^2}\approx area
\end{eqnarray}\\
It is interesting to note that for $r_0=0$ our results exactly reduces to the expression for entropy as 
obtained for R-N black hole\cite{Zhong} except the spin dependent term which is zero for the scalar field under consideration.
The result changes if the dilaton-axion black hole is extremal i.e. $r_+=r_-$. In that case,
\begin{eqnarray}
F_{extre}\approx\frac{-2\pi^3}{135\beta^4}\frac{(r_o+r_+)^{6n}}{(r_+-r_o)^{6n-6}\epsilon^3}
\end{eqnarray}
The expression for entropy for the extremal case therefore is,
\begin{eqnarray}
S_{extre}=\frac{8\pi^3}{135\beta^3}\frac{(r_o+r_+)^{6n}}{(r_+-r_o)^{6n-6}\epsilon^3}
\end{eqnarray}\\

It may be noted that in the extremal limit i.e. $r_+=r_-$, $\beta$ diverges ( see equ.19 ). However following
\cite{mitra} defining the cut-off $\epsilon$ in terms of proper radial variable, both the temperature as well as the 
entropy of the black hole vanishes\cite{wei,sgs}. It may further be noted that  
while for non-extremal case the entropy diverges linearly with $\epsilon$,
it diverges cubically for the extremal case.\\

Let us mention few results which are obtained for the coefficient 
of the logarithmic term through other methods. Solodukhin\cite{Solo} 
used path integral method as formulated by Gibbons and Hawking to 
find out the entropy of R-N black hole. The coefficient of the 
logarithmic term match nicely with the expressions as obtained by \cite{Zhong}for s=0.\\
It is interesting that for $r_0 = 0$ our result exactly matches with those obtained in these works if one puts $s=0$.  
In addition if we put $r_- = 0$, the magnitude of the 
coefficient of the logarithmic term will reduce to $\frac{1}{90}$ which is the correction for the Schwarzschild black hole. 

Furthermore for static spherically symmetric black holes the 
quantum entropies arising from fermionic 
field through brick wall model are $\frac{7}{2}$ times that of the 
scalar field and the entropy due to electromagnetic field is exactly twice of that 
for a scalar field.The spin of the fields in the expression of entropy 
depends on the rotation of the black hole or non-spherical symmmetry of the black hole\cite{jing}.\\

In other approaches, the principle of Quantum geometry yields the area law with 
logarithmic correction to the black hole entropy. From this approach   
the coefficient turns out to be $\frac{1}{2}$ instead $\frac{3}{2}$\cite{mitra2}.\\
For R-N black hole the magnitude of the coefficient obtained by quantum fluctuations appears as
$\frac{1}{2}$\cite{sou} while for BTZ black hole the corresponding term appears as $\frac{3}{2}$\cite{sou2}.\\

{{\Large{\bf 2.2.  Asymptotically  Non-Flat Dilaton-axion Black hole :}}}\\

We now extend our calculation for the asymptotically non-flat space-time. While for anti de-sitter and de-sitter space time
such extensions in the near horizon limit have been done in \cite{Winstanley, Kim}, 
here we extend it further for non-flat metric as discussed in \cite{Sur}.

We shall show later that our result will correctly reproduce the entropy expression in the flat space limit.  

The wave equation for asymptotically non-flat space-time is obtained as,
\begin{eqnarray}
\nonumber\frac{E^2f_{El}}{\frac{(r-r_+)(r-r_-)}{r^2(\frac{2r_o}{r})^{2n}}}+\frac{1}{r^2(\frac{2r_o}{r})^{2n}}
\nonumber\frac{\partial}{\partial r}[(r-r_+)(r-r_-)\frac{\partial
 f_{El}}{\partial r}]
-[\frac{l(l+1)}{r^2(\frac{2r_o}{r})^{2n}}+m^2]f_{El}=  o
\end{eqnarray} 
The r dependent radial wave number from this equation can be written as;
\begin{eqnarray}
k^2(r,l,E)=\frac{[E^2-\frac{(r-r_+)(r-r_-)}{r^2(\frac{2r_o}{r})^{2n}}(\frac{l(l+1)}{r^2(\frac{2r_o}{r})^{2n}}+m^2)]}
{(\frac{(r-r_+)(r-r_-)}{r^2(\frac{2r_o}{r})^{2n}})^2}
\end{eqnarray}\\
Only such values of E are to be considered here for which the above
expression is non-negative. Once again the radial degeneracy factor is obtained
through the following semi-classical quantisation condition.  
\begin{eqnarray}
 n_r\pi = \int_{r_h+\epsilon}^L dr k(r,l,E)
\end{eqnarray} 
where $n_r$ has to be a positive integer.\\
The expressions for free energy and entropy for asymptotically non-flat case can be calculated as
follows:
\begin{eqnarray}
\beta F_{boson} &=&\sum_{n_r,l,m_l}log(1-e^{-\beta E})
\nonumber \approx\int dl(2l+1)\int dn_r log(1-e^{-\beta E})\\
\nonumber && =-\int dl (2l+1)\int d(\beta E)\frac{n_r}{(e^{\beta 
E}-1)}\\
\nonumber &&  =-\frac{\beta}{\pi}\int dl (2l+1)\int \frac{dE}{(e^{\beta
 E}-1)}
\nonumber\int dr A(r)  \\
\nonumber &&  =-\frac{2\beta}{3\pi}\int_{r_h+\epsilon}^L
 dr\frac{r^2(\frac{2r_o}{r})^{2n}}
               {\Delta^2} \int_0^\infty dE B(r)\\  
\end{eqnarray}\\
where 
\begin{eqnarray}
 A(r) = 
\frac{[E^2-\frac{(r-r_+)(r-r_-)}{r^2(\frac{2r_o}{r})^{2n}}(\frac{l(l+1)}{r^2(\frac{2r_o}{r})^{2n}}
+m^2)]^\frac{1}{2}}
{\frac{(r-r_+)(r-r_-)}{r^2(\frac{2r_o}{r})^{2n}}}
\end{eqnarray}

\begin{eqnarray}
B(r)= 
=\frac{[E^2-\frac{(r-r_+)(r-r_-)}{r^2(\frac{2r_o}{r})^{2n}}m^2]^{3/2}}{(e^{\beta
 E}-1)}
\end{eqnarray}
and
\begin{eqnarray}
\Delta = \frac{(r-r_+)(r-r_-)}{r^2(\frac{2r_o}{r})^{2n}}
\end{eqnarray}
We will expand $\Delta$ about the event horizon similary as was done for asymptotically flat case.
Here the limits of integration is taken for values where the square
root is real. The  integration over $l$ is straightforward and has been explicitly carried out. Here also the $E$ integral  
can be evaluated only approximately.
The expression for entropy can be obtained from the formula,
\begin{eqnarray}
S_{boson} = \beta^2\frac{\partial F}{\partial \beta}
\nonumber  =\frac{1}{\epsilon}\frac{r_h^2(\frac{2r_o}{r_h})^{2n}}{360}\Delta_h'-[\frac{r_h^2(\frac{2r_o}{r_h})^{2n}}{360}(\frac{2\Delta_h'}{r_h (\frac{2r_o}{r_h})^{n}}-\Delta_h'')]
ln(\frac {L}{\epsilon})
\end{eqnarray}
where $\Delta_h'=\frac{(r_h-r_-)}{r_h^2}(\frac{2r_0}{r_h})^{2n}$ and $\Delta_h''=r_{h}^{-2}(\frac{2r_0}{r_h})^{-2n}-2(r_h-r_-)r_h^{-3}(\frac{2r_0}{r_h})^{-2n}+(r_h-r_-)r_h^{-2}(2r_0)^{-2n}2n r_h^{(2n-1)}+r^{-2}(\frac{2r_0}{r_h})^{-2n}-2r_h^{-3}(r_h-r_-)(\frac{2r_0}{r_h})^{-2n}+(r_h-r_-)r_h^{-2}(2n)(2r_0)^{-2n}r_h^{(2n-1)}$.

As in the previous case here also $r_h = r_+$.

It may be noted that for asymptotically non-flat case,if we put n=0, the metric becomes flat Reissner-Nordtrom metric
and the corresponding expression for the entropy matches exactly with the entropy of 
flat Reissner Nordstrom black hole as obtained in\cite{Zhong}.

The expression for $\beta$ for asymptotically non-flat metric ( see equ.3) is, 
\begin{eqnarray}
\beta = \frac{r_{+}^2(\frac{2r_o}{r_+})^{2n}}{(r_+-r_-)}
\end{eqnarray}
In this case the proper distance h corresponding to the ultraviolet cut-off $\epsilon$ is given by:
\begin{eqnarray}
\epsilon = \frac{p^2(r_+-r_-)}{4r_+^2(\frac{2r_o}{r_+})^{2n}}
\end{eqnarray}
Substituting the expression for $\beta$ and $\epsilon$; the expression for the entropy ( in the leading order ) becomes, 
Substituting the expression for  $\epsilon$; the expression for the entropy ( in the leading order ) becomes, 
\begin{eqnarray} 
S_{boson} = \frac{1}{90 p^2}r_+^2(\frac{2r_o}{r_+})^{2n}\approx area
\end{eqnarray}

{{\Large{\bf 3: Conclusion :}}}\\

By explicit and straightforward calculation we verify the entropy-area law for black hole for dilaton-axion coupled black holes
both for asymptotically flat and non-flat cases using the brick wall method.
Such black hole solutions are usually obtained from the low energy effective string action 
in presence of the dilaton and axion coupling. The results for the extremality condition have been derived. 
The entropy and temperature are shown to vanish in extremal limit.
Moreover the  leading order corrections to such black hole entropies for both asymptotically flat 
and non-flat dilaton-axion black hole turn out to be logarithmic in nature. The coefficients of the logarithmic term
have been determined in various cases and have been compared with those obtained from other methods.

\end{document}